\documentclass[fleqn,10pt]{olplainarticle}
\usepackage{url}
\pdfoutput=1
\title{Computational Thinking through Design Patterns in Video Games}

\author[1]{Giulio Barbero}
\author[2]{Marcello A. G{\'o}mez-Maureira}
\author[3]{Felienne F. J. Hermans}
\affil[1]{Leiden University, The Netherlands}
\affil[2]{Leiden University, The Netherlands}
\affil[3]{Leiden University, The Netherlands}


\begin{abstract}
Prior research has explored potential applications of video games in programming education to elicit computational thinking skills. However, existing approaches are often either too general, not taking into account the diversity of genres and mechanisms between video games, or too narrow, selecting tools that were specifically designed for educational purposes. In this paper we propose a more fundamental approach, defining beneficial connections between individual design patterns present in video games and computational thinking skills. We argue that video games have the capacity to elicit these skills and even to potentially train them. This could be an effective method to solidify a conceptual base which would make programming education more effective.
\end{abstract}

\begin{document}

\flushbottom
\maketitle

\section{Introduction}

Learning how to program involves more than absorbing syntax and semantics of a specific programming language, it also requires sensibility in combining and implementing these terms in an efficient and functional way. Programming makes use of procedural thinking, planning, data analysis, data re-elaboration and established praxes such as testing and debugging~\cite{Chng2019}. All these components and skills find definition under the concept of ``computational thinking''. Training computational thinking skills and being able to use them proficiently is a common objective for programming education and it is often one of the most challenging components for learners.

An important part of the research in the field is focused on finding new media and techniques to facilitate the development of these skills. Promising research has been conducted using computer games to train computational thinking components~\cite{wu2011facilitating}. 

Video games present advantageous characteristics for this scope: they can support problem-based learning, require information retrieval to succeed, provide immediate feedback allowing testing, can easily embed assessments and often create a social environment or community~\cite{Papastergiou2009}. They also motivate users with challenges and entertaining components~\cite{Prensky2003}. Prior research at the intersection of video games and computational skills has often been carried out in two main directions: one tends to embed and test these components in environments that were created specifically for that purpose, such as is generally the case in educational games~\cite{Weintrop2016}. The other seeks to analyse the effect of general gaming experiences (i.e., not purpose-built for personal improvement) on a set of computational thinking skills~\cite{Chng2019}.

While the first approach tends to deliver results for the study of the actual medium, the second takes a too general point of view that often faces noisy results due to the extreme diversity of elements in the game world. In this paper we argue that focusing on generalizable game features and design patterns that benefit the development of computational thinking skills offers a valuable middle-ground between these two approaches. We present this approach by outlining examples of design patterns that are most promising in the context of supporting programming education. 
In the following sections we present and describe the set of computational thinking skills we decided to use. We will then list and describe design patterns that we think can be positively connected to each of them also providing practical examples of where they are applied. The diversity and specificity of the examples suggests that each skill is activated by different video game components. Recognising these we can explore a new potential way to study the relation between gaming and computational thinking.

\section{Related Work}

\subsection{Computational Thinking and Programming Education}

The definition of computational thinking skills varies depending on the author, with different sets of overlapping components; often conceptually related to methods for data extraction and re-elaboration or logical and procedural reasoning. A commonly cited model comes from Kazimoglu et al.~\cite{Kazimoglu2012} and builds on the work of Wing 2006~\cite{wing2006computational}, Wing 2008~\cite{Wing2008}, Ater-Kranov et al.~\cite{ater2010developing} and Berland \& Lee~\cite{berland2011collaborative}. It lists five fundamental computational thinking skills. These are (1) conditional logic, (2) building algorithms, (3) debugging, (4) simulation and (5) distributed computation.

\textbf{Conditional logic} involves an understanding of true and false values and their use in control flow statements. This involves being able to evaluate the status of a system in a specific local statement and an understanding of how each operation manipulates it. \textbf{Building algorithms} is a form of step-by-step problem-solving that requires a more solution-driven view of the multiple conditional logic instances. It shares some overlap with the previous skill but in this case it is necessary to have an overview of all the single manipulations to understand how the system reaches a desired final status (i.e., the solution). \textbf{Debugging} describes the process of testing in order to spot and find solutions to problems in the code. \textbf{Simulation} refers to the creation of mental or physical models to define how to implement algorithms and which circumstances apply. Finally, \textbf{distributed computation} groups all the social aspects of programming, from project-oriented working to making use and contributing to a community~\cite{Kazimoglu2012}.

The main advantage of this set of computational thinking skills is its practicality; each skill is well-defined, easy to understand and covers a good part of the components that are necessary in the process of programming. It goes to depict a picture of computational thinking as a reasoning process that goes from the detail (conditional logic) to a larger view of the relations between them (building algorithms). It further includes practical elements that are necessary throughout the whole programming process such as debugging and simulating interactions between the components to reach the desired final state. Finally, programming is often an activity that heavily relies on the community behind it and distributed computation can be used to describe all the skills necessary to access, use and contribute to this community.

\subsection{Design Patterns in Video Games}
In order to identify useful video game components we follow the definition of ``game design patterns'' as described by Björk and Holopainen~\cite{bjork2004patterns}. Generally, design patterns are reusable structures for finding solutions to common problems in a domain (such as architecture~\cite{alexander1977pattern} or computer science~\cite{gamma1994DesignPatternsElements}). Depending on the field, this can range from the application of narrowly defined instructions to more general recommendations for specific circumstances. In the field of game research, Björk and Holopainen define game design patterns as \textit{``semi-formal interdependent descriptions of commonly reoccurring parts of the design of a game that concerns gameplay''}.

Game design patterns fit the purpose of our approach for multiple reasons. First, their definition is derived from a common term in the field of computer science. When studying computational thinking this is a beneficial connection, especially in terms of communication and structure of the knowledge for both the fields of computer science and game research. Second, they help to deconstruct video games into elements that can be studied and used more flexibly than focusing on the entirety of a game.

\section{Concepts of Computational Thinking in Video Games}

In this section we list and describe, for each computational thinking skill, some game design patterns that we consider useful for the activation and, perhaps, training of those skills.

\begin{figure}[th]
\centering
\includegraphics[width=\columnwidth]{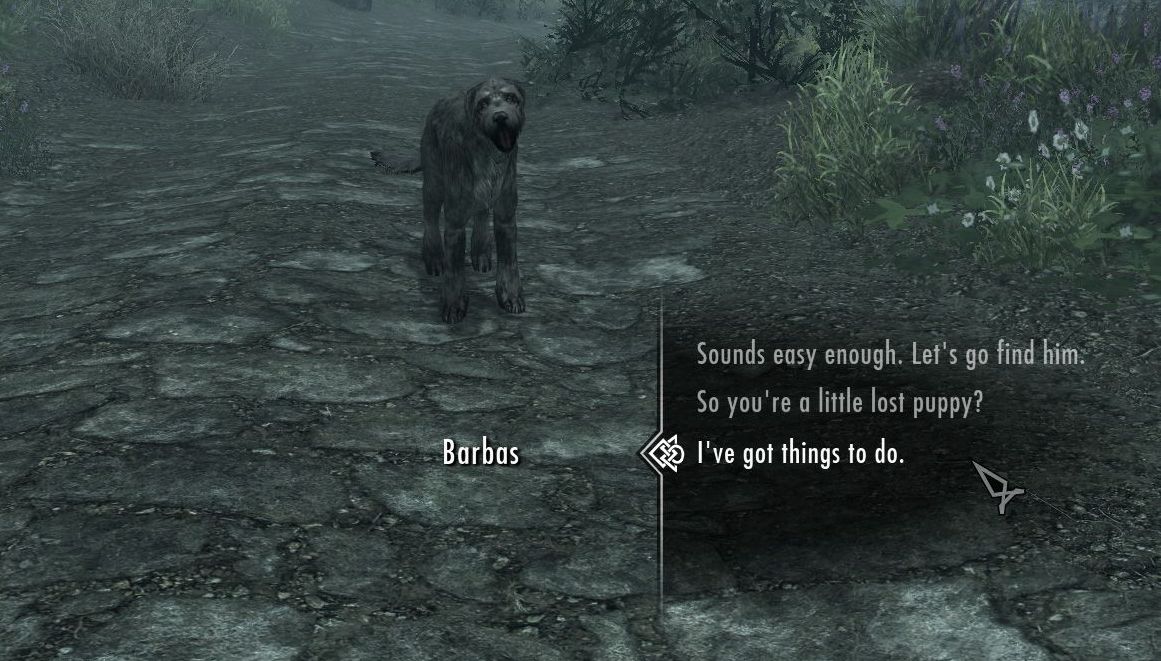}
\caption{Example of an interaction with mutually exclusive choices with a NPC in Skyrim}
\label{fig:skyrim}
\end{figure}

\textbf{Conditional logic}: Some could argue that most decisions, especially in video games, are binary, therefore based on conditional logic. Even though this could be the case, there are some particularly connected components in video games that are worth mentioning. In Bj\"{o}rk 2004 we find the pattern of \textbf{``Incompatible goals''} which refers to those situations in which pursuing a certain objective automatically forbids trying to pursue others. Players need to be able to evaluate the conditional status of those game elements that are triggering this pattern in order to understand the logical developing of the video game. Another pattern that requires the application of conditional logic is \textbf{``Varied gameplay''}. This pattern describes how certain choices and settings can provide the players with completely different game instances. Especially role-playing games (RPGs) serve as a fitting example, given that every decision players make opens up some paths while closing down others. A popular example can be found in the `The Elder Scrolls'~\cite{game:elderscrolls} series where different choices in the character creation and in the story itself lead to very different gameplay options and overall narrative experiences (see Figure~\ref{fig:skyrim}). This is facilitated by a sequence of conditional choices that allow and disallow certain features within the game as players progress.

\textbf{Building algorithms}: Building algorithms entails following a step-by-step plan to solve a problem. It requires the ability to individually evaluate those steps (using conditional logic) and to analyse the results of their sequential combination. Many of the game patterns that stimulate this skill make use of different aspects of this skill, requiring players to plan, concatenate and modulate the manipulations necessary to reach the desired status. A fitting game design pattern is the \textbf{``producer-consumer''} pattern which guides the use and importance of resources. In some games, it even determines the speed of the gameplay . In complex systems this pattern generates network of interrelated producers and consumers. Often, to reach a specific objective there are multiple steps of resources gathering, production and manipulation (which usually includes consumption) to be developed. Being able to foresee and plan over multiple cycles of discovering, extracting, transporting, storing and consuming resources requires similar mental mechanisms as building a computational algorithm. We can compare the producer-consumers to different functions returning elements as outputs and requiring outputs from other functions as input. The sequence of these elements and their inputs-outputs must be planned carefully in order to reach a certain goal. A very important concept of building an algorithm is taking a step-by-step approach~\cite{Kazimoglu2012} and, similarly, we can see a step-by-step approach when building a producer-consumer network in many `4X' games (a sub-genre of strategy games that involves eXploration, eXpansion, eXploitation, and eXtermination). This game design pattern is noted to conflict with the pattern ``predictable consequences'' which makes sense from a computational thinking point of view as well: complex algorithms with multiple steps and data manipulations are often more difficult to manage and usually require careful debugging.

Practical examples for this pattern are numerous, and it is arguably an essential component to most strategy games. For instance, in `Stellaris'~\cite{game:stellaris}, developed by Paradox Development Studio, certain jobs (tasks in the game) produce `minerals' that are then consumed to produce `consumer goods'. These can then be consumed again by other jobs to produce `research'. Since the ratio of consumption-production is hardly 1:1, this system requires players to carefully plan the construction of jobs, usually in order not to end up with a negative balance in any of the above mentioned resources. Another typical example is the complex and ramified network of resources of `Thea 2: The Shattering'~\cite{game:thea2}, developed by MuHa Games and Eerie Forest Studio. In this case we have four different tiers of resources with the last one being `crafted', consuming resources of the previous tier and requiring the acquisition of specific technologies. In this case we can see a sequence of steps necessary in order to acquire technologies, gather resources and craft them into a higher level one.

\begin{figure}[th]
\centering
\includegraphics[width=\columnwidth]{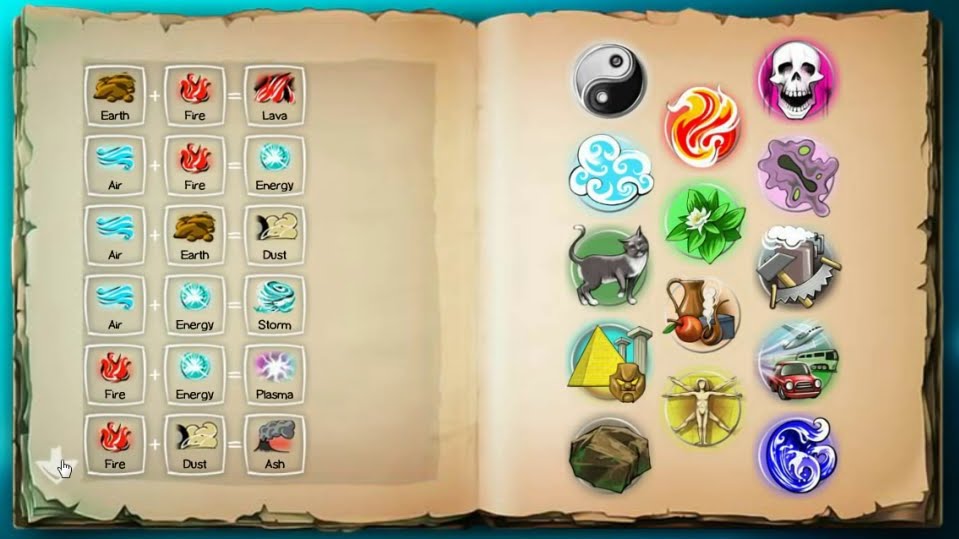}
\caption{Screenshot of `Doodle God' showing combinations of basic elements to create complex elements.}
\label{fig:doodlegod}
\end{figure}

\textbf{Debugging}: In order to analyse debugging we need to unpack the several elements that compose this skill. Debugging is understood as a process of trial and error that is developed through testing. A corresponding game design pattern is ``experimentation''. It usually indicates that a part of the game mechanics require a process of trial and error to be evident, understood or mastered. In the most extreme cases the whole gameplay revolves around experimentation in the form of `puzzles' to be solved. Experimentation is often realized through trial and error as well, with testing being a necessary component of it. Similarly, debugging usually involves being able to critically think about the current configuration and can necessitate multiple trials to determine where the errors are, as well as how to fix them efficiently. It is important to point out that experimentation is a quite broad pattern and its usefulness to debugging skills generally holds only under specific applications. If we want to better specify the context we need to limit it to the intersection with the game design pattern referred to as ``Puzzle-solving''. This refers to game features that need to be solved through inductive or deductive reasoning. If applied together with experimentation we are arguably defining an even closer reasoning to debugging; a mental process that, through deductive or inductive trials and errors attempts to spot and solve problems in the current solution. An interesting game example can be found in the mobile game `Doodle God'~\cite{game:doodlegod} by JoyBits (see Figure~\ref{fig:doodlegod}). In the game the player needs to combine basic elements (such as fire or water) to create new more complex ones (for example life or energy) that can be used to create even more complex elements. The whole game is based on a process of reasoning and, especially, trial and error. Players can think about potential element combinations and try them to see if they achieve new elements. The game also features helpful support for players that can partially (but rarely completely) provide hints to the creation of new components, highlighting one of the two elements that need to be combined.

\begin{figure}[th]
\centering
\includegraphics[width=\columnwidth]{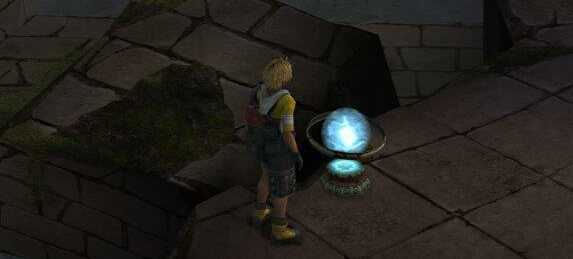}
\caption{Example of a saving point in Final Fantasy X}
\label{fig:finalfantasyx}
\end{figure}

\textbf{Simulation}: Similarly to `conditional logic', simulation is a very broad category that refers to essential concepts of many video games built as representations of real or imaginary phenomena. It seems self-evident that video games include simulations of some sort. However, it can be beneficial to focus on patterns that allow or elicit simulation skills within games themselves rather than understanding games as a simulation of real-life processes. Here we encounter some overlap with ``debugging'' since the game design pattern ``experimentation'' can once again be useful in this context. Players can create a set of possible actions and evaluate them using simulation skills. After, these actions are then validated by experimenting with them. In general, experimentation requires to activate simulation mental mechanisms in order to narrow the set of possibilities to try. Other game patterns do not directly trigger simulations but might favour it. Having ``Save-load cycles'' in a game is a typical example. With this pattern, players are allowed to save and load games at specific points (some games are more flexible on the saving locations), being able to replay certain challenges or actions. This can elicit simulation skills similarly to experimentation; players can simulate and select certain solutions and then try them in multiple rounds, loading back the game at every iteration. In the `Final Fantasy' game series (for instance `Final Fantasy X'~\cite{game:finalfantasyx}, see Figure~\ref{fig:finalfantasyx}), players can usually find saving spots right before the most challenging battles. In this way the player can try certain settings and, in case of failure, analyse their own errors improving on them after reloading the game from that last saving location.

\textbf{Distributed Computation}: We can identify several game design patterns that show useful traits to aspects of distributed computation. One example is ``communication channels'' which are present in many games that allow players to communicate with each other. ``Cooperation'' is another design pattern that can be reconnected to the idea of working together for a goal. However, we argue that it is even more compelling to notice how programming and gaming often behave similarly in their relationship with the respective communities. We would argue that that distributed computation is a skill that is necessary not only in computational thinking but also in many multiplayer games, or perhaps even certain single-player games with multi-agent aspects. Indeed both involve massive online communities interacting, debating and sharing information to complete tasks that would be hard or impossible to achieve alone. Moreover, even though both tend to also congregate on different platforms focusing on the context (e.g. GitHub for programmers and Steam for gamers),  sometimes these communities even interact on the same online networks such as Twitch or Reddit. The similarities do not stop here; as already mentioned, both the communities developed mainly online and became important resources for the fields (at least in many modern video games). In this sense, both communities based themselves on a ``Remix culture'' encouraging the sharing and re-elaboration of information, blurring the line between final consumer and contributor~\cite{ferguson2012everything, Madden2005TeenCC}. The overall argument in this case is that learning to make use and contribute to a gaming community probably involves similar skills to be able to do the same in a programming community because of these underlining similarities. A great example of an online community not directly connected to a video game (not referring to online gaming necessarily) is the community that formed around the `Elder Scrolls' series. Many online resources surrounding that series share information to both new and more seasoned players about how to develop their characters and how to customise their game experience. Another more general example is the massive amount of online videos of `playthroughs' (often referred to as ``Let's Play'' content) in which players record their game sessions while commenting and explaining their actions to show other players how to achieve a certain goal in a video game. Curiously we can find similar videos about programming, with (more or less) expert programmers coding and illustrating how to use certain languages, libraries or functions.

\section{Conclusions}
When highlighting video games as educative media for computational thinking purposes, we often tend to neglect the great variety of genres, mechanics and, more in general, game experiences. However, as we argue in this paper, the individual constituent game design patterns are perhaps a more fitting lens to assess the potential of a video game to improve computational thinking skills. Such individual elements can trigger very different thinking strategies and stimulate users in different ways. What we proposed is a different way of studying the connection between gaming and the development of programming skills, starting from the design elements that make up a video game rather than from the medium or specific game titles in general. This approach can also be instrumental to the creation of educational video games that make use of design concepts developed specifically for the medium rather than adapting them to the scope. In this paper we presented a examples of design patterns and games that could be investigated regarding their ability to improve computational thinking skills. Further research, both empirical and theoretical in nature, should focus on developing a design process to create or modify video games for that purpose. This would strengthen the case for the use of video games to improve computational thinking skills in general, and deepen our understanding for how to target such efforts towards individual skills.
\bibliography{sample.bib}

\end{document}